\RequirePackage{fix-cm}
\documentclass[twocolumn,epjc3]{svjour3}
\smartqed  
\RequirePackage{graphicx}
\RequirePackage[english]{babel}
\RequirePackage{amsmath} 
\RequirePackage{verbatim} 
\RequirePackage{mathrsfs}
\RequirePackage{amsfonts}
\RequirePackage{amssymb}
\RequirePackage[latin1]{inputenc}
\RequirePackage{hyperref}

\newcommand{\beq}{\begin{equation}}
\newcommand{\eeq}{\end{equation}}
\newcommand{\bea}{\begin{eqnarray}}
\newcommand{\eea}{\end{eqnarray}}
\newcommand{\nn}{\nonumber \\}
\newcommand{\nk}{\textbf{k}}
\newcommand{\dphi}{\delta \phi}
\newcommand{\x}{\textbf{x}}

\newcommand{\bra}{\langle}
\newcommand{\ket}{\rangle}
\newcommand{\mH}{\mathcal{H}}

\newcommand{\mR}{\mathcal{R}}
\newcommand{\Rh}{R_h = ct}

\journalname{Eur. Phys. J. C}

\begin{document}

  \title{Puzzling initial conditions in the $R_h=ct$ model}


\author{Gabriel R. Bengochea\thanksref{e1,addr1}
        \and
       Gabriel Le\'{o}n\thanksref{e2,addr2}
      }


\thankstext{e1}{e-mail: gabriel@iafe.uba.ar}
\thankstext{e2}{e-mail: gleon@df.uba.ar}

\institute{Instituto de Astronom\'\i
a y F\'\i sica del Espacio (IAFE), CONICET - Universidad de Buenos Aires, 1428 Buenos
Aires, Argentina \label{addr1}
\and
  Departamento de F\'{\i}sica, Facultad de Ciencias Exactas y
Naturales, Universidad de Buenos Aires, Ciudad Universitaria - Pab. I, 1428 Buenos
Aires, Argentina \label{addr2}
}

\date{Received: date / Accepted: date}

\maketitle

\begin{abstract}

In recent years, some studies have drawn attention to the lack of large-angle correlations
in the observed cosmic microwave background (CMB) temperature anisotropies with respect to
that predicted within the standard $\Lambda$CDM model. Lately, {it has been argued} that
such a lack of correlations could be explained in the framework of the so-called $R_h=ct$
model without inflation. {The aim of this work is to study whether there is
a mechanism} to generate, through a quantum field theory, the primordial power spectrum
presented by these authors. Specifically, we consider two different scenarios: first, we
assume a scalar field dominating the early Universe in the $R_h=ct$ cosmological model,
and second, we deal with the possibility of adding an early inflationary phase to the
mentioned model. {During the analysis of the consistency between the
predicted and observed amplitudes of the CMB temperature anisotropies in both scenarios,
  we run into deep issues which indicate that it is not clear how to characterize the
primordial quantum perturbations within the $R_h=ct$ model}.

\end{abstract}

\section{Introduction}\label{intro}

In addition to solve the horizon and flatness problems of the standard Big Bang model,
inflation generates a nearly scale invariant power spectrum for density perturbations,
which has been exquisitely tested with observations of the
cosmic microwave background (CMB) angular spectrum
\cite{wmap9,planck1,planck13inflation,planckBicep15,planckcmb2,planckinflation15}.

Starting with the Cosmic Background Explorer observations \cite{cobe}, it was noted
that the angular two-point correlation function at angular scales larger than 60 \(^\circ \)
is unexpectedly close to zero, contrary to what the standard $\Lambda$CDM model predicts.
Shortly after, it was rediscovered with the Wilkinson Microwave Anisotropy Probe (WMAP)
data \cite{wmap1} and later by Planck mission \cite{planck1,planck2}. This feature at
large scales was studied in detail by several authors, e.g.
\cite{copi1,copi2,copi3,sarkar,copi4}; it has been the source of some controversy (see
for instance \cite{efstat}) and today constitutes one of the persistent large-angle
anomalies in the CMB data \cite{copi5}.

Recently, a series of theoretical and observational motivations exposed in
\cite{F2,F3,F21} finished merging into what today is known as the $\Rh$ model \cite{F1}.
This model has received considerable attention over the last few years, since it has been
claimed to be favored over the standard $\Lambda$CDM by most observational
data \cite{F4,F5,F6,F7,F8,F9,F10,F11}. Even the authors argue that the mentioned horizon
problem could be solved in the framework of this model without an inflationary epoch at
the beginning of the Universe \cite{F19}. Basically, they hold that the Universe can be
described by a FLRW cosmology, where the cosmic fluid filling the Universe satisfies, at
all times, the overall equation of state $\rho+3P=0$, where $\rho$ and $P$ are the total
energy density and pressure of the cosmic fluid, respectively. According to the authors,
the condition $w=-1/3$ at all times is apparently required by the simultaneous application
of the Cosmological Principle and Weyl's postulate \cite{weyl,F20}. We remind the reader
that in the standard $\Lambda$CDM model, the equation of state $\rho+3P=0$ would lead to
a Universe with negative curvature.

However, some observational objections were raised, and also the validity of the physical
arguments underlying the $\Rh$ model have been criticized by a number of authors. Some of
them can be found for instance in \cite{OL1,OL2,miltra,BS1,shafer,OL3,L1,L2,kim1}.
In particular, the claim made in Ref. \cite{F17}, regarding that the
analysis of the CMB anisotropies in the $\Rh$ model is preferred over the $\Lambda$CDM,
appears to be incorrect; actually the formal computation of the angular power
spectrum, i.e. the $C_l$'s, is absent (in fact, in Sect. \ref{amplitud} we will
show explicitly that it is very unlikely that the $\Rh$ model can be made consistent with
the CMB observational data). Furthermore, the explanation, within this model, concerning
how $w$ is kept at $-1/3$ through the transitions from known matter to radiation sounds
at least questionable, and the idea that $\rho\propto a^{-2}$ throughout nucleosynthesis,
recombination, structure formation and today seems impossible to reconcile with all the
observations put together. Some of these criticisms are claimed to have been answered in
\cite{F12,F13,F14,F15,F16}. Nevertheless, after pointing out a number of objections to
the $\Rh$ model based on recent observational data, in Ref. \cite{kim2} the authors
analyzed the central assumption underlying the original theoretical argument for the
model, namely that the comoving Hubble radius should be constant, and showed that it is
not required.

In the present manuscript, we will focus on the results presented in
\cite{F17,F18}. There, the authors analyzed the CMB angular correlation
function for a fluctuation spectrum expected from growth in a Universe, whose dynamics is
constrained by the equation of state $w=-1/3$. To accomplish this, they mention that since
the exact form of the power spectrum emerging from the non-linear growth prior to
recombination is unknown, a parameterization for this spectrum can be performed, for
example, by assuming a scale-free initial power law spectrum and incorporating in its
shape other relevant effects. Then they ensure that it is possible to
obtain a better fit than the $\Lambda$CDM model to the data corresponding to the angular
correlation function, and conclude stating that the absence of power on large scales
exhibited by the angular correlation function might be evidence in support of the $\Rh$
model simply because it does not require inflation.

In this article, we perform a critical analysis whether there
might be a mechanism for generating, through a quantum field theory, the primordial
power spectrum presented by those authors in \cite{F17}. To do so, we are going to
consider two different scenarios: first, we will assume a scalar field dominating the
early Universe in the $R_h=ct$ cosmological model, and second, we will deal with the
possibility of adding an early inflationary phase to the mentioned model. After that, we
will analyze the consistency between the predicted and observed amplitudes of the CMB
temperature anisotropies in both scenarios.

The article is organized as follows: in Sect. \ref{classical}, we review some basics
about the $\Rh$ model and how to describe classical perturbations in that framework; in
Sect. \ref{quantum}, we search for a quantum mechanism to generate the primordial
curvature perturbation, and we obtain the primordial power spectra within $\Rh$ model with
and without an inflationary phase. Later, in Sect. \ref{amplitud}, we analyze the
amplitudes of the primordial power spectra and the consistency with the amplitude of the
CMB temperature anisotropies. In Sect. \ref{discusion} we make a discussion of our
results, and finally in Sect. \ref{conclusiones} we summarize our conclusions.

\section{Classical perturbations in the $\Rh$ model}\label{classical}

In this section, we provide a summary of the main characteristics of the $\Rh$ Universe.
Our main focus is the cosmological perturbations as presented in Refs. \cite{F1,F17,F18}.
The first two subsections will be heavily based on the results presented in those
references. However, in the last subsection, we will show how to relate the curvature
power
spectrum with the matter power spectrum proposed in Refs. \cite{F17,F18}.
We will use units in which $c=\hbar=1$. We will make use
of the reduced Planck mass $M_P^2 = 1/(8\pi G)$ and the ``West Coast'' signature $(+ -
- - )$ for the metric.

\subsection{The background}

The $\Rh$ Universe is characterized by a spatially flat FLRW spacetime, which in comoving
coordinates is represented by the line element $$ds^2 = dt^2 - a^2(t) \delta_{ij} dx^i
dx^j.$$
Additionally, the authors of the $\Rh$ Universe claim that the total matter
components in the Universe combined (dark matter, ordinary matter, radiation and dark
energy) behave as a perfect fluid with the overall equation of state\footnote{For the
specific
motivations behind the aforementioned equation of state, we refer the reader to the
original works by \cite{F1,F20} and a recent rebuttal regarding the consistency of the
physical motivations of the $\Rh$ Universe \cite{kim1,kim2}.}
\beq\label{eqedo}
P = -\frac{\rho}{3},
\eeq
where $\rho$ and $P$ represent the total energy density and pressure of the Universe,
respectively. Therefore, the Friedmann equation $H^2 \equiv (\dot a/a)^2 =  \rho/(3
M_P^2)$ and the continuity equation $\dot \rho + 3H (\rho + P) = 0$ (with the dot over
functions representing derivative with respect to cosmic time $t$) lead to a scale
factor of the form
\beq
a(t) = t/t_0,
\eeq
where we have normalized the scale factor to $a(t_0) = 1$ at the present
cosmic time $t_0$. Consequently, the Hubble radius evolves as $R_h \equiv H^{-1} = t$.
This is one of the main features of the $R_h$ model, i.e., the Hubble radius satisfies
the relation $H^{-1} = t$ during the whole cosmic evolution and not ``just today'' as in
the standard $\Lambda$CDM model. Therefore, the total energy density of the Universe
evolves as $\rho \propto 1/a^2$.

\subsection{Cosmological perturbations}

The dynamical evolution of the cosmological perturbations in the $\Rh$ Universe follows
from Einstein equations $\delta G_{ab} = \delta T_{ab} / M_P^2$. In particular, by
using the Newtonian (longitudinal) gauge, the
Fourier modes associated to the density contrasts defined as $\delta_{\nk} (t) \equiv
\delta \rho_{\nk} (t)/\bar \rho(t)$ (where $\bar \rho(t)$ is the background energy
density) satisfy
\bea\label{motiondelta0}
&\ddot \delta_k& + (2-w+3c_s^2 ) H \dot \delta_k \nn
&-& \frac{3}{2} H^2 (1+8w-3w^2-6c_s^2)
\delta_k =  - \frac{k^2 c_s^2}{a^2} \delta_k.
\eea
Therefore, by considering the equation of state associated to the $\Rh$ Universe, it is
assumed
that $w = c_s^2 = -1/3$. Thus, the motion equation for $\delta_k$ is given by
\beq\label{motiondelta}
\ddot \delta_k   + \frac{3}{t}    \dot \delta_k - \frac{1}{3} \frac{\Delta_k^2}{t^2}
\delta_k  = 0
\eeq
where $\Delta_k \equiv k/(aH)$. Note that $aH= H_0$ is a constant ($H_0$
denotes the Hubble parameter today).
It is not hard to find the solutions of \eqref{motiondelta}; nevertheless, the solutions
depend on whether $k$ is greater or less than $aH$, i.e. $\Delta_k > 1$ or $\Delta_k <
1$. If $\Delta_k > 1$, then the solutions are a growing mode  $\delta_k \sim t^{\alpha}$
(with $\alpha > 0$)  and a decaying mode. On the contrary, if $\Delta_k < 1$, then the
solutions are  a constant and a decaying mode. In the $\Rh$ Universe one is
primarily interested in the modes such that $k > aH$ since these are the modes that can
grow into the large scale structure. As a matter of fact,  motivated by the angular
correlation of the CMB and the solution corresponding to the growing modes of Eq.
\eqref{motiondelta}, the authors of Refs. \cite{F17,F18} proposed that
the initial matter power spectrum is of the form
\beq\label{PSfulvio0}
P_\delta (k) \propto k - b \bigg( \frac{2\pi}{R_e(t_e)}  \bigg)^2 \frac{1}{k}
\eeq
where $b$ is an unknown constant to be adjusted, and $R_e$ is the proper distance to the
last
scattering surface at time $t_e$, which corresponds to the cosmic time at the decoupling
epoch. The power spectrum \eqref{PSfulvio0} can be recast as
\beq\label{PSfulvio2}
P_\delta(k) = A  \mH \bigg[ \frac{k}{\mH} - b \bigg( \frac{\theta_{\textrm{max}}}{a(t_e)}
 \bigg)^2 \frac{\mH}{k} \bigg]
\eeq
where $A$ is the amplitude of the power spectrum, $\mH \equiv a H$ and
$\theta_{\textrm{max}}$ is the maximum angular size of any fluctuation associated with
the CMB  emitted at $t_e$, that is, $\theta_{\textrm{max}} = [2 \pi
a(t_e)]/[k_{\textrm{max}} R_e(t_e)]$; also $k_{\textrm{max}} /\mH=1$.

\subsection{Curvature and matter power spectra in the $\Rh$
Universe}\label{curvatureperts}

Our next step is to relate, also through a classical analysis, the matter power spectrum
with the curvature power spectrum in
the $\Rh$ Universe. Later, we will investigate whether it is possible to find a quantum
mechanism for
generating the curvature perturbation. If possible, we will relate that spectrum with the
matter power spectrum, and then we will compare it with the one proposed in
\eqref{PSfulvio2}.

We start the discussion by switching to conformal time $\eta$, i.e. $dt^2 = a^2 d\eta^2$.
In these coordinates $\mH \equiv aH = a'(\eta)/a(\eta)$, where a prime denotes
derivative with respect to conformal time. As a matter of fact, using the equation
of state $P = -\rho/3$, the continuity equation $\rho'+3\mH(\rho + P) =0$ and
Friedmann equation $\mH^2 = a^2 \rho / 3 M_P^2$, we arrive at the important result
\beq\label{cfhubble}
\mH = H_0.
\eeq
That is, $\mH$ is a constant of motion in the $\Rh$ Universe and has the value of the
Hubble parameter today. For the sake of completeness, we present the explicit form of the
scale factor in conformal time coordinates:
\beq\label{factorescala}
a(\eta) = e^{H_0(\eta-\eta_0)}
\eeq
where $\eta_0$ corresponds to the conformal time today.

The most generic metric associated to a flat FLRW Universe with linear scalar
perturbations is
\bea\label{metricagenerica}
ds^2 &=& a^2(\eta) \{  (1-2\varphi) d\eta^2 + 2(\partial_i B) dx^i d\eta  \nn
&-& [ (1-2\psi)\delta_{ij} + 2 \partial_i \partial_j E     ]dx^i dx^j \},
\eea
where $\varphi,\psi,E,B$ are scalar functions of the spacetime. In the Newtonian gauge,
$\varphi = \Phi$, $\psi = \Psi$ and $E=B=0$.

In the absence of anisotropic stress, Einstein equations (EE) $\delta G_{ab} = \delta
T_{ab} / M_P^2$ lead to $\Phi = \Psi$. Moreover, considering once again that in the
$\Rh$ Universe $c_s^2 = w  = -1/3$, the equation of motion for the Fourier mode $\Phi_k(\eta)$ 
that results from combining EE is
\beq\label{motionPhifulvio}
\Phi_k'' + 2\mH \Phi_k' - \frac{k^2}{3} \Phi_k = 0.
\eeq
The general solution of the former equation is a linear combination of
$\exp[(q-\mH)\eta]$ and $\exp[-(q+\mH)\eta]$, with $q\equiv+\sqrt{k^2/3 + \mH^2
}$. Furthermore, using Eqs. \eqref{cfhubble} and \eqref{factorescala}, we can
express the scale factor as $a(\eta) \propto \exp (\mH\eta)$. Consequently, if $k \ll \mH$
then $q \sim \mH$, thus the linearly independent solutions of \eqref{motionPhifulvio} can
be approximated by a constant and a decaying mode $\exp(-2\mH\eta) \propto a(\eta)^{-2}$.
On the other hand, if $k \gg \mH$ then $q\sim k/\sqrt{3}$, and the linearly
independent solutions of \eqref{motionPhifulvio} are approximately given by a growing
mode $\Phi_k \sim \exp[(k-\mH)\eta]$ and a decaying mode $\exp[-(k+\mH)\eta] \propto
\exp(-k\eta)/a(\eta)$ (note that the conformal time $\eta$ is an increasing variable).

The EE with component $\delta
G_{00} = \delta T_{00} / M_P^2$ is useful to relate the density contrasts with the metric
perturbation $\Phi$. That is,
\beq\label{deltagenerica}
\delta_{k} = -\frac{2}{3} \frac{k^2}{\mH^2}  \Phi_k - 2 \Phi_k -\frac{2}{\mH} \Phi_k'.
\eeq

As we mentioned in the previous subsection, in the $\Rh$ Universe one is interested in
the modes such that $k>\mH$; that is, the modes whose associated proper wavelength is
less than the Hubble radius. These are the modes that evolve as $\Phi_k \sim
\exp{(k-\mH)\eta}$; consequently $\Phi_k' = (k-\mH) \Phi$. By using that result, Eq. 
\eqref{deltagenerica} becomes
\beq\label{deltafulvio}
\delta_k = -\frac{2}{3} \frac{k^2}{\mH^2}  \Phi_k  - 2 \frac{k}{\mH} \Phi_k
\eeq
We emphasize that Eq. \eqref{deltafulvio} is valid only for $k > \mH$ and $w = c_s^2 =
-1/3$.

At this point we have to do a technical digression. The quantum analysis of the field
perturbations usually involves the so called Mukhanov-Sasaki variable, and then one
relates that variable with the comoving curvature perturbation $\mR$. We will follow such
an analysis in the next section; however, Eq. \eqref{deltafulvio}, which will help us to
relate the matter power spectrum with the curvature one, was obtained in the Newtonian
gauge. Therefore, it will be useful to change from the Newtonian gauge to the comoving
gauge. That relation is generically given (for constant $w$) by \cite{mukbook}
\beq\label{RPhi}
\mR = -\frac{5+3w}{3+3w} \Phi - \frac{2}{3+3w} \mH^{-1} \Phi'.
\eeq
Thus, for $w=-1/3$, Eq. \eqref{RPhi} leads to $\mR = -2 \Phi - \mH^{-1} \Phi'$. Moreover,
if we
focus on the modes such that $k>\mH$ and recall that for such modes $\Phi'_{k} = (k-\mH)
\Phi_k$, then we arrive at
\beq\label{RPhifulvio}
-\mR_k  = \bigg(1 + \frac{k}{\mH} \bigg) \Phi_k.
\eeq

With Eqs. \eqref{deltafulvio} and \eqref{RPhifulvio} at hand, it is straightforward (in
the comoving gauge) to
relate the corresponding matter and curvature spectra, namely
\beq\label{eqchingona}
P_\delta(k) \simeq \frac{4}{9} P_{\mR} (k) \bigg( \frac{k^2}{\mH^2} +  4 \frac{k}{\mH} -2
\bigg),
\eeq
where we have retained only the first three dominant terms in powers of $k/\mH$.

Equation \eqref{eqchingona} is the main result of this subsection. One can immediately
observe that if $P_{\mR} \propto k^{-1}$, then the resulting matter power spectrum will be
of a similar structure as the one shown in Eq. \eqref{PSfulvio2}, except for a constant
term.

In the following section, we will attempt to construct a mechanism for generating the
curvature
power spectrum $P_{\mR}(k)$.

\section{Generation of the primordial curvature perturbation}\label{quantum}

In this section, we will consider two possibilities for generating the primordial
curvature perturbations: a scalar field dominating the early $\Rh$ Universe, and a
preceding inflationary era in the $\Rh$ Universe.

Since in the $\Rh$ model the combination of different types of matter is such
that it mimics a perfect fluid with an overall equation of state $P=-\rho/3$ (which
involves
a negative pressure), we will make the standard assumption that the early Universe was
dominated by a scalar field $\phi(\x,t)$, with some potential $V(\phi)$, such
that $P$ and $\rho$ associated to $\phi$ satisfy $P(\phi)=-\rho(\phi)/3$ at all times.
Afterwards, the scalar field should decay into particles of the standard model and
possibly into dark matter particles, and the evolution of the Universe then follow the
$\Rh$ model.
Since we are considering a canonical scalar field, the action is given by
\beq
S[\phi] = \int d^4 x \sqrt{-g} \bigg( \frac{1}{2} g^{ab} \nabla_a \phi \nabla_b \phi -
V(\phi)      \bigg)
\eeq

In contrast with the standard $\Lambda$CDM model (plus inflation) in which the end of a
different cosmological era is linked to a change in the equation of state, in the $\Rh$
Universe the equation of state $P=-\rho/3$ should be satisfied at all times during the
evolution of the Universe. As a consequence, we need to provide a condition that
marks the end of the early cosmological era dominated by the field $\phi$. We propose
that the value of the adiabatic speed of sound  $c_s^2$ will help to
provide such condition.

For ordinary matter and constant equation of state we know that $w=c_s^2$. However,
for a scalar field generically $c_s^2 \neq w$. In particular, for a canonical
scalar field (a field with canonical kinetic term), $c_s^2=1$ \cite{mukbook}. In fact, in
standard slow-roll inflation $c_s^2 = 1$ and $w \simeq -1$. Therefore, in the $\Rh$
Universe, we will consider a canonical scalar field that dominates the matter content
of the early Universe, and such scalar field will be characterized by $c_s^2=1$. Then, at
some point during the evolution, the scalar field will decay in such a way that $c_s^2$
will decrease from $c_s^2=1$ to $c_s^2 = {-1}/{3}$. Note from Eq. \eqref{motiondelta0}
that it is crucial to have $c_s^2 = w = -1/3$ in order to obtain Eq. \eqref{motiondelta}, which
results in a
solution for the growing modes. It is important to mention that other combinations of $w$
and $c_s^2$ would lead to a solution of Eq. \eqref{motiondelta0} with a growing mode;
in particular, the condition for the $\Rh$ model is $w =-1/3$. Hence, other values of
$c_s^2$, but maintaining $w=-1/3$ could lead to a growing mode in the $\Rh$ model. On the
other hand, Eq. \eqref{motiondelta} is the main equation used by the authors of the
$\Rh$ model to analyze the growth of structure in \cite{F17,F18}; and to
obtain Eq. \eqref{motiondelta} from Eq. \eqref{motiondelta0}, one must satisfy $c_s^2 = w
= -1/3$.

To continue, we split the scalar field into an homogeneous part plus small
inhomogeneities, i.e.
$\phi(\x,t) = \phi_0(t) + \dphi(\x,t)$. The homogeneous part of the field drives the
background evolution, that is, the one characterized by the $\Rh$ Universe, and the
quantum theory of $\dphi(\x,t)$ will result in the primordial power spectrum of the
perturbations. In the following, we will attempt to construct a quantum theory for
$\dphi$, but first we will derive some useful quantities to describe the background.

Since the background field, $\phi_0$, drives the evolution of the $\Rh$ Universe, we can
associate the standard energy-momentum tensor $T^\alpha_\beta$ to the field $\phi_0$. In
particular, from the time component $T^0_0 = \rho(\phi)$, we infer $\rho = \phi_0'^2/2a^2
+ V(\phi)$; additionally, $T^i_j = -P(\phi) \delta^i_j$ implies that $P = \phi_0'^2/2a^2
- V(\phi)$.

Using the fact that $\mH$ is constant [see \eqref{cfhubble}], and from the continuity
equation
$\rho' + 3\mH (\rho + P)=0$ applied to the scalar field $\phi_0$, one obtains
\beq\label{phicero}
\phi_0'' = 0.
\eeq

Consequently, from Eqs. \eqref{cfhubble} and \eqref{phicero}, it is clear that in the
$\Rh$ Universe, $\mH$ and $\phi_0'$ are exactly constants of motion. As a matter of fact,
using the Friedmann equations, it can be shown that
\beq\label{eqchida}
\frac{\phi_0'}{\mH} = \sqrt{2} M_P.
\eeq

Furthermore, using that $\phi_0''=0$ and that $\mH$ is a constant, we can find a potential
that is consistent with $P(\phi)=-\rho(\phi)/3$. This potential turns out to be
\beq
V(\phi) = \mH^4 e^{-2\phi/\mH}.
\eeq

Now, let us focus on the linear scalar perturbations. The field perturbations $\dphi$
induce metric perturbations $\delta g_{\mu \nu}$ via EE.
As we mentioned in Sect. \ref{curvatureperts}, Eq. \eqref{metricagenerica} represents
the most generic metric associated to a FLRW Universe with
scalar perturbations. As is well known, the relativistic perturbation theory has the
issue of gauge redundance \cite{mukhanov1992,jmartin}. However, the gravitational part can
be characterized
by a single, gauge-invariant object known as the Bardeen potential defined as
\cite{bardeen}
\beq
\Phi_B (\x,\eta) = \varphi + \frac{1}{a} [a(B-E')]'.
\eeq
In the same manner, the matter sector can be described by the gauge-invariant field
perturbation
\beq
\dphi^{(\textrm{gi})} (\x,\eta) = \dphi + \phi_0'(B-E').
\eeq

The Einstein equations relate $\Phi_B$ and $\dphi^{(\textrm{gi})}$ through a constraint
equation. That implies that the scalar sector can be characterized by a single object;
this object is the so-called Mukhanov-Sasaki variable, defined by
\beq
v(\eta,\x) \equiv a \left[\dphi^{(\textrm{gi})} + \phi_0' \frac{\Phi_B}{\mH}  \right]
\eeq
All other relevant quantities can be expressed in terms of $v(\eta,\x)$, i.e. it fully
characterizes the scalar sector.

Moreover, we can expand the action of our theory, that is, the action of a scalar
field
minimally coupled to gravity, up to second order in the scalar perturbations, obtaining
\beq\label{accion1}
\delta S^{(2)} = \frac{1}{2} \int d\eta d^3x \left[ (v')^2 - \delta^{ij} \partial_i v
\partial_j v  + \frac{z''}{z} v^2   \right],
\eeq
where $z \equiv a \phi_0'/\mH$. From Eq. \eqref{eqchida} we obtain that, in the $\Rh$
model
\beq\label{zeta}
z = \sqrt{2} M_P\: a
\eeq

From the action \eqref{accion1}, the equation of motion is
\beq
v'' - \nabla^2 v - \frac{z''}{z} v = 0.
\eeq
Notice that Eq. \eqref{zeta} implies that  $z''/z = a''/a$. Additionally, the fact that
$\mH'=0$
implies that $a''/a = a'^2/a^2 = \mH^2$. Thus, the equation of motion can be rewritten as
\beq\label{motionv}
\partial^2 v - \mH^2 v= 0,
\eeq
where we have defined the operator $\partial^2 \equiv \partial_\eta^2-\nabla^2$. Since
$\mH^2$ is a positive constant, Eq. \eqref{motionv} is a Klein-Gordon type of equation
with the
``wrong'' mass sign, that is, the motion equation of a free tachyon field. This can
also be read directly from action \eqref{accion1}, which is $\delta S^{(2)} =  \int
d\eta d^3 x \mathcal{L}$, where
\beq\label{lagr}
\mathcal{L} = \frac{1}{2} \partial^2 v + \frac{1}{2} \mH^2 v^2.
\eeq
Thus, quantizing the scalar field $v(\eta,\x)$ in the $\Rh$ Universe, is equivalent of
quantizing a free tachyon  with constant mass given by $m^2=-\mH^2<0$.

There are various methods proposed for constructing a quantum field theory of a
free tachyon in the past \cite{bose,argentinos,tanaka,feinberg,AS,DS}. Nevertheless, there
are some issues that seem to
be always present in such theories \cite{kamoi}. Among them, we can mention the
non-locality of the tachyonic
field, represented in the present paper by the field $v(x)$ [the short-hand notation $x$
refers to a point in spacetime $(\x,\eta)$], in the sense that the commutator (as well as
the anti-commutator in some methods) $[\hat{v}(x),\hat{v}^{\dag}(x')]$ does not vanish
for spacelike arguments. Another puzzle is that the energy operator, normally associated
to the Hamiltonian, does not have a lower bound on its spectrum, i.e. there are
infinitely negative energy states, which requires some reinterpretation principle
\cite{bilaniuk}. But perhaps
the most serious difficulty in formulating a theory of tachyons is that the resulting
$S$-matrix is non-unitary. Thus, it is unknown how to describe interactions within the
theory of a tachyonic field \cite{kamoi}.

In spite of the aforementioned issues, we could proceed in a pragmatic way, and construct
a quantum theory
of the field $v(x)$ but only considering the modes such that $k > \mH $, i.e. modes with a
proper wavelength
less than the Hubble radius $\lambda_p < H^{-1}$. Also, according to Ref. \cite{F17} those
modes are the ones
that can grow and evolve into large scale structure.\footnote{Note that modes with
$k\ll\mH$ are always less than $\mH$ in the $\Rh$ model, therefore they will not be
relevant at all observationally.} Afterwards, we could compute the quantum two-point
correlation function and
extract its corresponding power spectrum.

There are various known methods for constructing a quantum theory for
a field with the Lagrangian \eqref{lagr} that ignores the ``problematic modes''. Among
those, we can mention the one proposed by Feinberg \cite{feinberg} and another one
developed by
Arons and Sudarshan (AS) \cite{AS}. We will focus on those methods as they illustrate the
kind of
puzzles one encounters when trying to compute $\bra \hat v(\x,\eta) \hat v (\x',\eta)
\ket$.

Both methods assume that the field $v(x)$ possesses non-vanishing Fourier components
only for $k \geq \mH$ and is expanded as
\bea
v(x) &=& \frac{1}{(2\pi)^{3/2}} \int_{k \geq \mH} \frac{d^3k}{\sqrt{2w(k)}} \: [
c_{+} (\nk) e^{-iw(k)|\eta| + i \nk \cdot \x} \nn
&+& c_{-} (\nk) e^{iw(k)|\eta| - i \nk
\cdot \x}],
\eea
where $w(k) \equiv+ \sqrt{ k^2 - \mH^2}$. Then one promotes the field $v$ into an
operator $\hat v$. The difference between the AS and Feinberg's method is the
quantum interpretation of the coefficients  $c_{+} (\nk)$ and $c_{-} (\nk)$.

Feinberg's method follows the traditional approach of promoting
$c_{+} (\nk) \to \hat c (\nk) $ and $c_{-} (\nk) \to \hat c(\nk)^{\dag}$ into
annihilation and creation operators respectively.  Furthermore, $\hat c(\nk)$  and
$\hat c(\nk)^{\dag}$ satisfy anti-commutator relations $\{\hat c(\nk), \hat
c(\nk')^{\dag}   \} = \delta^3 (\nk-\nk') $. The anti-commutator replaces the commutator
since the former is compatible with Lorentz invariance, within the quantization of a
free tachyon. However, under a suitable Lorentz transformation, $\hat c(\nk)$ can be
converted into $\hat c(\nk)^{\dag}$. Thus, the vacuum state defined as $\hat
c(\nk) | 0 \ket = 0$ is not an invariant vacuum state since in another frame of reference
it takes the form $\hat c(\nk)^{\dag} |0 \ket = 0$. For this reason, we find Feinberg's
method not to be suitable for the problem at hand.

On the other hand, in the AS method \textit{both coefficients are promoted to
annihilation operators}.  The fact that both operators $\hat c_{+}
(\nk)$ and $\hat c_{-} (\nk)$ are annihilation operators is needed in this approach in
order to preserve the Lorentz invariance symmetry of the vacuum state
\cite{bose,kamoi}. Moreover, one also has anti-commutation relations $\{\hat c_{\pm} (\nk),
\hat  c_{\pm}(\nk')^{\dag}   \} = \delta^3 (\nk-\nk') $ and the vacuum state defined as
$c_{\pm} (\nk) |0 \ket$. Consequently, we can compute $\bra 0|\hat v(\x,\eta) \hat
v^{\dag}
(\x',\eta)|0 \ket$, which yields
\beq\label{2ptos}
\bra 0|\hat v(\x,\eta) \hat v^{\dag} (\x',\eta)|0 \ket = \int_0^\infty \frac{dk}{k} \:
\frac{\sin k|\x-\x'|}{k|\x-\x'|} \frac{k^3}{2\pi^2w(k)}.
\eeq

In the comoving gauge, the curvature perturbation is given by $\mR = v/z$. That is, from
\eqref{2ptos} we can extract the primordial power spectrum $P_{\mR}(k,\eta) =
{P_v(k)}/{z(\eta)^2}$ which, using Eq. \eqref{zeta}, results in
\beq\label{PSfulvio}
P_{\mR}(k,\eta) = \frac{1}{2M_P^2 a^2(\eta) w(k)} \simeq
\frac{1}{2M_P^2 a^2(\eta) k} .
\eeq
The previous approximated expression is valid only for $k > \mH$. As a matter of fact,
$P_{\mR}(k,\eta)
=  0$ for $k< \mH$; i.e. there are no ``super-Hubble'' modes (see footnote 2).

Substituting Eq. \eqref{PSfulvio} into Eq. \eqref{eqchingona} yields the matter
power spectrum,
\beq\label{PSmatterfulvio}
P_\delta (k,\eta)
\simeq  \frac{2}{9 M_P^2 a^2(\eta) \mH  }   \bigg(\frac{k}{\mH} +4 - 2 \frac{\mH}{k},
\bigg)
\eeq
which is valid for $k>\mH$, while $P_\delta(k) = 0$ if $k < \mH$.
The quantum theory proposed above resulted in  a matter power spectrum
\eqref{PSmatterfulvio} of the same structure in $k$, plus a constant term, as the one in
Eq. \eqref{PSfulvio2}, whose
form was proposed by the authors of \cite{F17,F18} motivated by observational data.
It may be the case that the spectrum \eqref{PSmatterfulvio}, including the constant term,
could reproduce the results obtained from the one proposed heuristically in Refs.
\cite{F17,F18},  Eq. \eqref{PSfulvio2},  for some values of the parameters considered in
those references. However,  the quantum theory of the primordial perturbation in the
present
section contains at least two fundamental issues: (i) The theory describes a free
tachyon field and (ii) the final primordial spectrum, Eq. \eqref{PSfulvio}, depends on
the scale factor. We will study the implications of the second issue in the next section.
Here, let us focus on the first issue.

The fact that the spectrum obtained involved the quantum theory of a free tachyon field
could discourage some readers to consider the quantum theory of the field $v(x)$ as a
serious mechanism for generating the primordial spectrum in the $\Rh$ Universe. The
reasons are vast and we entirely subscribe to most of them. However, a possible way to
deal with that issue is to
abandon the $\Rh$ model framework for the early Universe and instead use the standard
inflationary paradigm. In other words, we can assume that inflation did occur in the
early Universe, but then, after the reheating era, the Universe followed the evolution
described by the $\Rh$ Universe.

In slow-roll inflation, one has the standard theory of the inflaton field, and the end of
the inflationary era is achieved when the
slow-roll parameters are close to unity. As is well known, the quantum theory of inflation
leads to the following expression for the Mukhanov-Sasaki variable:
\beq
v_k(\eta) \simeq \frac{1}{\sqrt{2k}} \bigg(1-\frac{i}{k\eta} \bigg) e^{-ik\eta},
\eeq
and $z=\sqrt{2\epsilon} M_P \:a$, where $\epsilon$ is the standard Hubble slow-roll
parameter defined as $\epsilon \equiv 1-\mH'/\mH$ and during inflation $\epsilon \ll 1$.
As a
consequence, the primordial spectrum for slow-roll inflation is $P_\mR \simeq
|v_k(\eta)|^2/z^2$, that is,
\beq
P_{\mR} (k) \simeq \frac{1}{4 M_P^2 \epsilon a^2 k } \bigg( 1 + \frac{\mH^2}{k^2} \bigg),
\eeq
where we used that $\mH \simeq -1/\eta$ during inflation.
For the ``super-Hubble'' modes, i.e. modes that $k \ll \mH$ during inflation, one has the
familiar result (ignoring the numerical factors)
\beq
P_{\mR} (k) \simeq \frac{H^2}{ M_P^2 \epsilon k^3 },
\eeq
that is, the scale invariant primordial spectrum, which remains constant after
the ``horizon crossing.'' On the other hand, the ``sub-Hubble'' modes, which satisfy $k
\gg \mH$ during inflation, lead to a spectrum of the form
\beq\label{PSinflationsubH}
P_{\mR} (k) \simeq \frac{1}{M_P^2 a^2 \epsilon k}.
\eeq

$ $ $ $ In the standard $\Lambda$CDM model, the sub-Hubble modes are ignored since they
decay as $\sim a^{-2}$. However, in the $\Rh$ Universe these modes need not to
decay so much because in the $\Rh$ model there is no ``horizon
problem'' (see Ref. \cite{F19}). Consequently, there is no minimum value of
$e$-folds needed for inflation to solve the horizon problem. Thus, the inflationary era
could end after a few $e$-folds and the sub-Hubble modes still contribute to the
observable modes in the CMB.

Additionally, in the inflation plus $\Rh$ Universe scenario, the dynamical evolution
of $\mH$, ignoring the reheating era, is
\begin{eqnarray}\label{Hhibrido}
\mH
= \begin{cases} -\frac{1}{\eta}
\:\: \textrm{if $\eta < \eta_f$ (inflation)}\\
H_0 \:\: \textrm{if $\eta \geq \eta_f$ (the $\Rh$ Universe)},  \\
\end{cases}
\end{eqnarray}
where $\eta_f$ denotes the conformal time at which inflation ends, thus, $\eta_f =
-1/H_0$.

$ $ Equation \eqref{Hhibrido} implies that super-Hubble modes $k<\mH$
during inflation, stay super-Hubble at all times, namely they do not re-enter the
horizon as in the standard $\Lambda$CDM model, instead, if a mode satisfies $k < \mH$
during inflation, then it also satisfies $k < \mH$ during the whole evolution of the
$\Rh$ Universe. Therefore, taking into account that $\Phi_k$ is a constant for
super-Hubble modes and that $w=-1/3$,  one obtains from Eqs.  \eqref{deltagenerica} and
\eqref{RPhi} the matter power spectrum,
\beq
P_\delta(k) \simeq \frac{H_*^2}{M_P^2 \epsilon_* k^3},
\eeq
where $H_*$ and $\epsilon_*$ are valuated at the time $-k\eta_* = 1$ during inflation.

On the other hand, Eq. \eqref{Hhibrido}  implies that if a mode is sub-Hubble
during the $\Rh$ Universe expansion, $k > \mH = H_0$, then it is also sub-Hubble during
inflation $-k\eta > 1$ (or equivalently $k > \mH_{\textrm{inf}}$). As a consequence,
the primordial spectrum associated to these modes, Eq. \eqref{PSinflationsubH}, should be
evaluated at some conformal time $\tilde \eta$ after inflation ends, namely when
$\epsilon = 1$ and $a(\tilde \eta) > a(\eta_f)$. Therefore, after substituting Eq.
\eqref{PSinflationsubH} into Eq. \eqref{eqchingona}, the matter power spectrum associated
to the super-Hubble modes, at leading order in $k/\mH$, is
\beq
P_\delta(k) \simeq \frac{k}{M_P^2 a^2(\tilde \eta) \mH}
\eeq

Thus, adding a standard inflationary era to the $\Rh$ model, results in a matter
power spectrum of the form
\begin{eqnarray}\label{PSmatterhibrido}
P_\delta (k)
\propto \begin{cases} k
\:\: \textrm{if $k > \mH = H_0 $ }\\
k^{-3} \:\: \textrm{if $k < \mH = H_0$ }.  \\
\end{cases}
\end{eqnarray}

The spectrum \eqref{PSmatterhibrido} has a resemblance  to the standard prediction of the
$\Lambda$CDM model but with an
important difference. In the traditional model, one divides the parts proportional to
$k$ and $k^{-3}$ using the value $k_{\textrm{eq.}}$, instead of the value $H_0$ as in
\eqref{PSmatterhibrido}, where we have defined $k_{\textrm{eq.}} \equiv
1/\eta_{\textrm{eq.}}$ and
$\eta_{\textrm{eq.}}$ denotes the conformal time at the epoch of matter-radiation
equality. If $k \ll k_{\textrm{eq.}}$, then the mode enters the horizon (or becomes
sub-Hubble) during the matter dominated epoch and $P_\delta (k) \propto k$. On the
contrary, if  $k \gg k_{\textrm{eq.}}$, then the modes becomes sub-Hubble during the
radiation era and $P_\delta (k) \propto k^{-3}$.

Also, note that the matter power spectrum shown in \eqref{PSmatterhibrido} is not
of the form
proposed by the authors of the $\Rh$ Universe [see \eqref{PSfulvio2}], which we obtained
by adding a previous inflationary phase to the $\Rh$ model. The only similarity between
the two expressions, Eqs. \eqref{PSmatterhibrido} and \eqref{PSfulvio2}, is in the
term that goes as $k$, the rest of the terms are not equivalent. The spectrum of Eq.
\eqref{PSfulvio2} was  proposed heuristically (not deduced from a physical mechanism) by
the original authors of the $\Rh$ model in order to provide a solution for the low
correlation observed at large angles. Given that Eq. \eqref{PSmatterhibrido} is not the
same as \eqref{PSfulvio2}, we cannot say if the analysis made by the authors of $\Rh$
model still is valid for Eq. \eqref{PSmatterhibrido}, i.e. we cannot claim that the
spectrum \eqref{PSmatterhibrido} solves the low large-angle correlation. In the next
section, we will deepen the discussion regarding the viability of
the primordial spectra obtained when considering the amplitude of the temperature
anisotropies in the CMB.

\section{Amplitude of the primordial spectra and the CMB temperature
anisotropies}\label{amplitud}

In the previous section, we proposed a mechanism for deriving the primordial spectrum
from the quantum fluctuations of a field $\phi$ that dominated the early Universe, but
with
the condition that the field must satisfy $P(\phi) =-\rho(\phi)/3$. That
procedure resulted, albeit the need of a quantum theory for a free tachyon, in a
prediction with a similar shape to the one proposed by the authors of the $\Rh$ Universe
\cite{F17,F18}, but with
the difference that the final primordial spectrum showed in Eq. \eqref{PSfulvio} depends
on the scale factor. In the present section,
we return to this subject by analyzing the amplitude of the spectrum, which is tightly
related to the amplitude of the CMB
temperature anisotropies.

We will consider the Sachs-Wolfe effect on the temperature anisotropies. That
effect is the dominant source for the anisotropies at large angular scales ($l \leq 20$).
It also relates the anisotropies in the temperature observed today on the celestial
sphere to the inhomogeneities in the Newtonian potential on the last scattering surface,
\beq
\frac{\delta T}{T} (\theta,\varphi) \simeq \frac{1}{3} \Phi (\eta_D,\x_D).
\eeq
Here, $\eta_D$ is the conformal time of the decoupling era and $\x_D = R_D (\sin \theta
\sin \varphi, \sin \theta \cos \varphi, \cos \theta)$, with $R_D$ the comoving radius of
the last scattering surface. It is useful to
perform a multipolar series expansion $\frac{\delta T}{T} = \sum_{l,m} a_{lm} Y_{lm}
(\theta,\varphi)$.

Using the Fourier expansion of $\Phi(\eta_D,\x_D)$ and the
expression for the Sachs-Wolfe effect, the coefficients $a_{lm}$ can be expressed as
\beq\label{alm}
a_{lm} \simeq \frac{4\pi i^l}{3} \int \: \frac{d^3k}{(2\pi)^{3/2}} j_l (kR_D) Y_{lm}^*
(\hat k) \Phi_k (\eta_D),
\eeq
with $j_l$ the spherical Bessel functions of order $l$. The observed data is presented in
terms of the angular power spectrum defined as $C_l \equiv 1/(2l+1) \sum_m |a_{lm}|^2$,
that is, Eq. \eqref{alm} yields
\beq\label{cl}
C_l \simeq \frac{2}{9\pi} \int_0^\infty dk\: k^2 j_l (kR_D)^2 |\Phi_k (\eta_D)|^2.
\eeq
It is straightforward to check that if $|\Phi_k(\eta_D)|^2 = A/k^3$, with $A$ some
constant, then $C_l = A/[l(l+1)]$. In other words, for $l \leq 20$, the quantity
$l(l+1)C_l$ is a constant $A$ and is equal to the amplitude
of the squared temperature anisotropies $\sim 10^{-9}$ \cite{planckcmb2}. Thus, it is a
necessary condition
that the Newtonian potential at the time of decoupling should scale as
$\sim k^{-3}$ if $C_l$ is to be consistent with the temperature anisotropies of
the CMB.

In the standard $\Lambda$CDM model, the value of $|\Phi_k(\eta_D)|^2$ is determined by
the modes that became super-Hubble during inflation. Those modes behave as
$|\Phi_k^{\textrm{inf}}|^2 = A/k^3$ and remained constant during the whole cosmological
evolution up until they became sub-Hubble at some point. If the modes became
sub-Hubble during the matter dominated epoch, then they remain constant even for $k>\mH$.
On the other hand, the modes that became sub-Hubble during the radiation dominated epoch
decayed as $\sim a^2$. Thus, in the traditional scenario once $|\Phi_k^{\textrm{inf}}|^2$
is generated during inflation, it remains fixed at that value and then one simply relates
$|\Phi_k(\eta_D)|^2 \propto |\Phi_k^{\textrm{inf}}|^2$. In other words, the amplitude $A=
k^3 |\Phi_k^{\textrm{inf}}|^2$ is fixed during inflation and is the same for all modes up
to
the decoupling epoch.

Now, let us focus on the value of $ |\Phi_k (\eta_D)|^2$ in the $\Rh$ Universe. As
mentioned previously, the potential $\Phi_k$ corresponds to the general solution of Eq.
\eqref{motionPhifulvio}, which  is a linear combination of
$\exp[(q-\mH)\eta]$ and $\exp[-(q+\mH)\eta]$, with $q\equiv+\sqrt{k^2/3 + \mH^2
}$. Moreover, using $a(\eta) \propto (\exp
\mH\eta)$, the two linearly independent solutions can be rewritten as follows: the first
solution is $\exp[(q-\mH)\eta] = \exp[(q/\mH-1)\mH\eta] = \exp[(\alpha-1) \mH \eta]
\propto a^{\alpha-1}$. We have defined $\alpha \equiv q/\mH$; similarly, the second
solution is given by $a^{-\alpha-1}$. Since $\alpha > 0$ the second solution
corresponds to a decaying mode; on the other hand, the first solution is explicitly
\beq\label{phifulvio}
\Phi_k (\eta)  = C_k a(\eta)^{\alpha-1}.
\eeq
If $k < \mH$ then $\alpha \sim 1$; on the contrary, if $k > \mH$ then $\alpha > 1$.
Therefore, depending on whether $k< \mH$ or $k>\mH$, the first linearly
independent solution of  \eqref{motionPhifulvio} can be approximated by a constant or a
growing mode [which is consistent with the discussion after Eq. \eqref{motionPhifulvio}].
In the $\Rh$ model one is interested in the growing mode, hence $k > \mH$ and $\alpha >
1$.

The primordial spectrum  $P_\mR(k)$ obtained in Eq. \eqref{PSfulvio} can
be related to the  amplitude of the Newtonian potential $|\Phi_k|^2$
through Eq. \eqref{RPhifulvio}, which results in
\beq\label{phiprimordial}
|\Phi_k (\eta_p)|^2 = \frac{\mH^2}{2M_P^2 a_p^2 k^3}.
\eeq
Note that we have evaluated the scale factor, and consequently the power spectrum, at
some conformal time $\eta_p$, i.e. $a(\eta_p) = a_p$.

At this point, we will make the assumption that the value of $|\Phi_k|^2$ obtained during
the period dominated by the scalar field $\phi$ [Eq. \eqref{phiprimordial}], when $c_s^2 =
1$, is the same as the one given by \eqref{phifulvio}, when $c_s^2 = -1/3$ at the time
$\eta_p$. Note, however, that in both cases $w=-1/3$, hence the $\Rh$ Universe expansion
remains unchanged. In particular, we are assuming that the following condition is
satisfied:
\beq\label{condition}
|\Phi_k (\eta_p)|^2_{c_s^2 = 1} = |\Phi_k (\eta_p)|^2_{c_s^2 = -1/3}
\eeq
but $w=-1/3$ in both situations. In other words, we are assuming that the
``reheating'' period  in the $\Rh$ Universe is practically instantaneous.

Furthermore, with the condition \eqref{condition} and expression \eqref{phiprimordial}, we
can find the explicit value of the integration constant $C_k$ in \eqref{phifulvio}:
\beq
C_k^2 = \frac{\mH^2}{2 M_P^2 k^{3} a_p^{2\alpha}}.
\eeq
Consequently, the expression for $|\Phi_k(\eta)|^2$ in the $\Rh$ model is given by
\beq\label{potencialnewtfulvio}
|\Phi_k(\eta)|^2 =  \frac{\mH^2}{2M_P^2 k^3 a_p^{2\alpha}}
a(\eta)^{2\alpha-2}.
\eeq

The Newtonian potential obtained, Eq. \eqref{potencialnewtfulvio}, has a scale
dependence $k^{-3}$, which could guarantee the same amplitude for all the modes. But,
unfortunately, it also carries an additional $k$-dependence through $\alpha$;
specifically, for modes
 $k > \mH$, we can approximate $\alpha \simeq k/\mH$. Therefore, different modes, grow at
a different rate, and for that reason, the amplitude of each mode at the time of
decoupling would be different for each mode. Nevertheless, we could make use of the fact
that up to this point $a_p$ has remained unspecified. Then, to avoid the mentioned issue,
we must have $a_p^{2 \alpha} =
N^2 a_D^{2\alpha-2}$, with $N^2$  some normalization constant and $a_D$ the scale factor
at the time of decoupling. In other words, we are adjusting the value of $a_p$ for each
mode in order to achieve that all modes arrive with the same amplitude at the time of
decoupling, and thus we obtain a nearly scale invariant spectrum as observed in the CMB.

By using the expression $a_p^{2\alpha} = N^2 a_D^{2\alpha-2}$ and Eq.
\eqref{potencialnewtfulvio}, we obtain the value of $|\Phi_k|^2$ evaluated at the time of
decoupling,
\beq
|\Phi_k(\eta_D)|^2 =  \frac{\mH^2}{2M_P^2 k^3 N^2}.
\eeq
Afterwards, we could simply adjust $N^2$ so  that

$$\mH^2/(M_P^2 N^2) \simeq 10^{-9}.$$
However,
the condition $a_p^{2\alpha} = N^2 a_D^{2\alpha-2}$ can be rewritten as
\beq\label{condicionchingona}
\log {a_p} = \frac{1}{\alpha} \log N + \bigg( 1 - \frac{1}{\alpha} \bigg) \log a_D,
\eeq
which implies that if $ \alpha \gg 1$ (or equivalently $k \gg \mH $) then $a_p \simeq
a_D$. That is, for these modes, the epoch dominated by the scalar field
$\phi$ should last up until the decoupling epoch in order to the primordial spectrum
obtained can
have a consistent amplitude with the corresponding observed in the temperature
anisotropies of the CMB.

Another way to show that the spectrum \eqref{potencialnewtfulvio} presents some issues
with the CMB is to calculate the angular power spectrum $l(l+1)C_l$ from Eq. \eqref{cl}
using precisely Eq. \eqref{potencialnewtfulvio}. It is known that, for low $l$, say $l<20$,
the shape of the angular power spectrum must be essentially a constant, independent of
$l$, which results from a nearly scale invariant primordial power spectrum. That
is, the region of the angular spectrum where the Sachs-Wolfe effect is dominant must not
depend on $l$ in order to be consistent with the CMB. Thus, our next task will be to
compute the angular power spectrum using the spectrum \eqref{potencialnewtfulvio} for
$l<20$.

Assuming that $a_p = a_D/\gamma$, with $\gamma > 1$ a constant and evaluating Eq.
\eqref{potencialnewtfulvio}  at the time of decoupling, we have
\begin{equation}\label{potencialdecoupling}
  |\Phi_k(\eta_D)|^2 =  \frac{\mH^2}{2M_P^2 a_D^2} \frac{\gamma^{2\alpha}}{k^3}.
\end{equation}
Substituting Eq. \eqref{potencialdecoupling} into Eq. \eqref{cl}, we find
\begin{equation}\label{integralfulvio}
  C_l \simeq \frac{1}{9\pi} \frac{\mH^2}{M_P^2 a_D^2} \int_{H_0}^{\infty} \frac{dk}{k}
 j_l (kR_D)^2 \gamma^{2\alpha}.
\end{equation}
Since the spectrum \eqref{potencialnewtfulvio} was obtained for the modes $k > \mH = H_0$ note that the lower
limit of integration is $H_0$. As a matter of
fact, the modes with $k < \mH$ vanished when we considered the quantum theory of a free
tachyon. The value of $R_D$, which corresponds to the comoving radius of
the last scattering surface, was calculated in Refs. \cite{F17,F18} in the context of the
$\Rh$ model resulting $R_D \simeq 10/H_0$. Additionally, for $k>\mH$ we can
approximate $\alpha \simeq k/\mH = k/H_0$. Performing the change of variable in the
integral \eqref{integralfulvio} $x \equiv k/H_0$ yields
\begin{equation}\label{integralchingona}
   C_l \simeq \frac{1}{9\pi} \frac{\mH^2}{M_P^2 a_D^2} \int_{1}^{\infty} \frac{dx}{x}
 j_l (10x)^2 \gamma^{2x}.
\end{equation}
Noting that the asymptotic form of the functions $j_l(10x)^2$ contribute with a
factor of $1/x^2$ as $x \to \infty$ and that $\gamma > 1$, we can conclude that the
integral in Eq. \eqref{integralchingona} diverges.

We remind the reader that the standard $\Lambda$CDM
prediction, which is consistent with the CMB data, would have resulted in  $l(l+1)
C_l = $ const. for the lowest multipoles (approximately for $l<20$). On the other
hand,
the result of Eq. \eqref{integralchingona} diverges even for the lowest values of $l$. 
Thus, in addition to the
aforementioned problems, e.g. the need of a quantum theory of a free tachyon and that the epoch dominated by the scalar field
$\phi$ should last up until the decoupling epoch,  the resulting angular
power spectrum is divergent.

Furthermore, the condition \eqref{condicionchingona} also applies to the situation in
which
one drops the $\Rh$ model in the early Universe in favor of an inflationary era. As we
described in Sect. \ref{quantum}, during inflation, the dynamical evolution of the
Mukhanov-Sasaki variable $v$ leads to the following expression for the comoving
curvature perturbation:
\beq
|\mR_k(\eta)|^2 \simeq \frac{1}{4M_P^2 \epsilon a^2 k} \bigg( 1+ \frac{\mH^2}{k^2} \bigg),
\eeq
which for sub-Hubble modes $k>\mH$ is approximated by $|\mR_k(\eta)|^2 \simeq (4M_P^2
\epsilon a^2 k)^{-1}$. Next, we evaluate $|\mR_k(\eta)|^2$ at some time $\eta_p$
near the end of inflation, i.e.  when $\epsilon =1$, and we make the assumption that
\beq
|\mR_k(\eta_p)|^2_{w\simeq-1} = |\mR_k(\eta_p)|^2_{w = -1/3},
\eeq
once again neglecting the reheating era.

$ $ $ $ Using Eq. \eqref{RPhifulvio}, which allows us to relate
$|\mR_k(\eta)|^2$ with $|\Phi(\eta)|^2$ when $w=-1/3$, and since
we are connecting the inflationary regime with the $\Rh$ Universe expansion, we have 
$\mH =
H_0$.  Consequently, the amplitude of the primordial Newtonian potential is
\beq\label{phinflprimordial}
|\Phi(\eta_p)|^2 \simeq \frac{\mH^2}{4 M_P^2 k^3  a_p^2},
\eeq
which is essentially the same as the one in \eqref{phiprimordial}. Therefore, all the
mathematical steps that lead from \eqref{phiprimordial} up to \eqref{condicionchingona}
are equivalent, including  Eq. \eqref{potencialnewtfulvio}. And, from condition
\eqref{condicionchingona}, one is led to conclude that inflation should
last
until the decoupling epoch in order to the primordial spectrum obtained can have a
consistent
amplitude with the temperature anisotropies of the CMB. Moreover, the discussion
regarding the shape of the angular power spectrum also remains the same since the
spectrum obtained in Eq. \eqref{potencialnewtfulvio} will be exactly the same in the
present scenario of adding an inflationary era to the $\Rh$ model.

Given that both approaches, for generating the primordial perturbation, require
very unlikely conditions to be compatible with the observed shape and amplitude of the
temperature anisotropies, we could do a search for the initial value of
$|\Phi_k(\eta_p)|^2$
so that it is consistent with the CMB temperature anisotropies based solely in the
dynamics
of the $\Rh$ model.

The equation of motion for $\Phi_k$, Eq. \eqref{motionPhifulvio}, implies that
\beq
\frac{d}{d\eta}\bigg( \Phi_k(\eta) a(\eta)^{1-\alpha}   \bigg) = 0,
\eeq
That is, $ \Phi_k(\eta) a(\eta)^{1-\alpha} $ is a constant of motion in the $\Rh$
Universe. Consequently, we have the relation $\Phi_k(\eta_D) = ({\Phi_k
(\eta_p)}/{a_p^{\alpha-1}} ) a_D^{\alpha-1}$, which implies that
\beq\label{amplitudphigenerica}
|\Phi_k(\eta_D)|^2 = \frac{|\Phi_k
(\eta_p)|^2}{a_p^{2\alpha-2}}  a_D^{2\alpha-2}
\eeq
We now select the value of the scale factor at the initial time $\eta_p$. Hence, we
assume that $a_P = C a_D$, with $C$ some normalization constant and $C<1$, which from Eq.
\eqref{amplitudphigenerica} yields
\beq\label{amplitudphigenerica2}
|\Phi_k(\eta_D)|^2 = \frac{|\Phi_k(\eta_p)|^2}{C^{2\alpha-2}}.
\eeq

Therefore, if the initial amplitude of the Newtonian potential is of the form
\beq\label{phiprimordialfulvio}
|\Phi_k(\eta_p)|^2 = \frac{C^{2\alpha}}{k^3},
\eeq
then the amplitude of the Newtonian potential at the time of decoupling, Eq.
\eqref{amplitudphigenerica2}, is $ |\Phi_k(\eta_D)|^2 = C^2/k^3$, which will be consistent
with the amplitude of the temperature anisotropies if $C^2 \simeq 10^{-9}$.

It is evident that Eq. \eqref{phiprimordialfulvio} is not equivalent to 
\eqref{phiprimordial} and/or \eqref{phinflprimordial}, which corresponds to the
primordial amplitude obtained in the two approaches described in the previous section.
Thus, any mechanism proposed for generating the primordial curvature perturbation in the
$\Rh$ Universe must be of the form of Eq. \eqref{phiprimordialfulvio} in order to be
consistent with the CMB temperature anisotropies. However, neither a scalar field
dominating the early Universe satisfying an equation of state $P(\phi) = -\rho(\phi)/3$
nor the inflaton yield a primordial spectrum compatible with Eq.
\eqref{phiprimordialfulvio}.

\section{Discussion}\label{discusion}

In this work, {we began by proposing} that the small inhomogeneities of a
scalar field $\dphi$ could generate
the primordial spectrum in the same fashion as the one in the standard inflationary
scenario, but with the important difference that during the period dominated by the
scalar field, the equation of state $P(\phi) = -\rho(\phi)/3$ should be satisfied at all
times. That is an important condition in the $\Rh$ cosmological model.

Under that proposal, the quantum theory of perturbations led to a theory of a free
tachyon field with mass
$m^2 = -\mH^2 = -H_0^2 < 0$. We pushed forward and followed a suitable method for dealing
with that kind of theory, which resulted in a matter power spectrum \eqref{PSmatterfulvio}
that is similar in structure in $k$, plus a constant term, to the one proposed in an
empirical manner in Refs. \cite{F17,F18}, Eq. \eqref{PSfulvio2}. It might be the case that,
for some values of the parameters corresponding to the spectrum proposed by the authors of
the Refs. \cite{F17,F18}, the spectra \eqref{PSmatterfulvio} and \eqref{PSfulvio2}
coincide and the analysis of Refs. \cite{F17,F18} continues to be valid for the matter
power spectrum given in Eq. \eqref{PSmatterfulvio}. Nevertheless, as we will see in the
rest of this discussion there are other important problems associated to the spectrum,
Eq. \eqref{PSmatterfulvio}.

The fact that the quantum theory of the perturbations in the $\Rh$ Universe
resulted in that of a free tachyon carries deep issues. Among them, perhaps the most
important issue in the cosmological context is that the corresponding
$S$-matrix is non-unitary. Therefore, there is no clear way how to describe interactions.
The interactions with other fields are important since at some point the scalar field
$\phi$, dominating the early Universe in the $\Rh$ model, should decay into the particles
of the Standard Model, and in the absence of a well defined $S$-matrix, it is a puzzle how
to describe such interactions. Another related issue with the non-unitarity of the
$S$-matrix is that the self-interactions of the scalar degree of freedom, characterized
here by the Mukhanov-Sasaki variable, result in primordial non-Gaussianities; however,
since there is no way to characterize the interactions, one cannot quantify the amount of
primordial non-Gaussianities generated by the perturbations in the $\Rh$ Universe.

Since in the $\Rh$ framework $z''/z = a''/a$, the action given in Eq. \eqref{accion1}
will be identical
for a hypothetical analysis of tensor modes. Thus, if one wanted to study the tensor case,
the quantum field theory used here for scalar perturbations would be equivalent.
Therefore,
in the light of our results, it is not obvious how tensor modes could be generated
within the framework of a quantum theory in the $\Rh$ model. In case of adding an
inflationary
epoch prior to the $\Rh$ evolution, tensor modes could be generated but since the modes
satisfying $k>\mH$ are relevant, their amplitudes will be exponentially suppressed
and the suppressing will continue also during the $\Rh$ evolution.

Those issues added to the technical and conceptual problems raised by a quantum theory of
a
free tachyon might suggest that we should abandon the idea of describing the quantum
perturbations in the $\Rh$ Universe.

The aforementioned problems led us to consider the quantum theory of perturbations
during an inflationary era preceding the $\Rh$ cosmological expansion. However, given the
nature of
the $\Rh$ Universe, we needed to focus on the primordial spectrum of the sub-Hubble
modes. The primordial spectrum that resulted from inflation for the sub-Hubble modes is
not consistent with the matter power spectrum proposed by the authors of the $\Rh$
Universe.

In fact, the matter power spectrum that we obtained by adding an early inflationary
regime in the $\Rh$ model, resembles to
the traditional one from the $\Lambda$CDM model, but it is not exactly the same. More
precisely, the matter power spectrum obtained by adding an early inflationary regime to
the $\Rh$ model,  can be separated into two cases. In one case the spectrum goes as $k$,
and in the second case the spectrum goes as $k^{-3}$. In the $\Lambda$CDM model the
matter power spectrum can also be separated into two cases, one that goes as $k$ and a
second case where the spectrum goes as $k^{-3}$. However, as mentioned after
Eq. \eqref{PSmatterhibrido}, the conditions for the separation into two cases in the
$\Rh$ model are not the same as the conditions in the standard model. Consequently, the
functional form of the $\Lambda$CDM matter power spectrum is completely different from the
$\Rh$ model. Moreover, the matter power spectrum obtained by adding an
early era of inflation to the $\Rh$ model leads to a matter power spectrum that is
different from the one proposed empirically by the authors of the $\Rh$ model. The main
motivation, as stated by those authors when proposing such a spectrum, was to solve the
observed low correlation at large angles in the angular correlation, and since the
spectrum that we have obtained, Eq. \eqref{PSmatterhibrido}, is not equal to the one
heuristically proposed, Eq. \eqref{PSfulvio2}, we cannot claim that the spectrum
in Eq. \eqref{PSmatterhibrido} explains the observed low angular correlation at large
angles.

Finally, we investigated the predicted amplitude of the CMB temperature anisotropies
following the two approaches described. In the first approach, we considered an early
phase dominated by a scalar field $\phi$ satisfying the equation of state of the $\Rh$
model; in the second framework, we assumed an inflationary stage preceding
the $\Rh$ cosmological evolution. In both approaches, the amplitude of the Newtonian
potential at the time of decoupling, which is the main source of the temperature
anisotropies at low angular multipoles, depends on the wavenumber $k$ in a non-trivial
way. The reason is that in the $\Rh$ Universe the evolution of each mode associated to
the Newtonian potential evolves as $\sim a^{\alpha-1}$, with $\alpha \simeq k/\mH$. As a
consequence, we were forced to choose a particular initial condition for the evolution of
the modes that translates into adjusting the value of the scale factor for each mode at
the initial time $\eta_p$. Specifically, we had to choose $a_p^{2\alpha} = N^2
a_D^{2\alpha-2}$, with $N$ some normalization constant and $a_D$ the value of the scale
factor at the time of decoupling. Such an election implies that, for modes $k \gg \mH$,
the initial value of the scale factor is $a_p \simeq a_D$. That is, the era dominated by
the scalar field $\phi$, in the first approach, or the inflationary era preceding the
$\Rh$ evolution, should last up to the decoupling epoch in order to the primordial
spectrum obtained
can have a consistent amplitude with the corresponding observed in the temperature
anisotropies of the CMB. Another related  problem that emerges from considering the
Newtonian potential Eq. \eqref{potencialnewtfulvio}   is that the predicted angular power
spectrum Eq. \eqref{integralchingona} diverges in the region where the Sachs-Wolfe effect
is dominant. The expected behavior for the angular power spectrum, which is consistent
with the CMB data in the Sachs-Wolfe region, is $l(l+1)C_l \simeq$ constant.

We ended our analysis by obtaining the desired form of the initial amplitude of the
Newtonian potential based solely on the dynamics of the $\Rh$ Universe, and that it could
be
consistent with the CMB temperature anisotropies. That primordial amplitude should be
$|\Phi_k(\eta_p)|^2 = C^{2\alpha}/k^3$, which is not the one obtained from the two
approaches considered so far. Thus, we think that any proposed mechanism for generating
the primordial spectrum should predict that particular initial amplitude to be consistent
with the
observed CMB anisotropies. In fact, if some
physically motivated mechanism, within the $\Rh$ model, can reproduce the primordial
spectrum $|\Phi_k(\eta_p)|^2 = C^{2\alpha}/k^3$ then all the concerns raised in our paper
would possibly disappear. Nevertheless, neither the inflaton nor a scalar field
dominating
the $\Rh$ Universe can produce that kind of spectrum.

\section{Conclusions}
\label{conclusiones}

Some studies have drawn attention to the lack of large-angle correlations in the observed
CMB temperature anisotropies with respect to that predicted within the standard
$\Lambda$CDM model. Recently, some authors have suggested that this
lack of correlations could be explained in the framework of the so-called $R_h=ct$ model
without inflation, by selecting an explicit form for the matter power spectrum and showing
that it could achieve a better fit than the $\Lambda$CDM model to the data corresponding
to the CMB angular correlation function. The aim of this work was to
critically investigate whether there may be a mechanism to generate, through a quantum
field theory, the primordial power spectrum presented by these authors.

During this search we run into deep issues and, we also studied the possibility of
adding an inflationary phase prior to the evolution given by the mentioned model. The
resulting power spectrum for the relevant sub-horizon modes within this approach is not
consistent with the matter power spectrum displayed by the mentioned authors; thus, it cannot 
explain the unexpectedly close to zero angular two-point correlation function observed
at angular scales larger than 60\(^\circ \).

Also, we analyze the consistency between the predicted and observed amplitudes of the CMB
temperature anisotropies with and without the inflationary epoch added prior to the $\Rh$
evolution. We found that for modes satisfying $k\gg\mH$, the epoch dominated by the scalar
field (representing the matter field in the $\Rh$ Universe or the
inflaton) should last up to the decoupling epoch in order to the primordial spectrum
obtained can have a consistent amplitude with the corresponding observed in the
temperature anisotropies of the CMB. That is an implausible condition. Additionally, we
have performed a brief analysis by focusing on the lowest angular multipoles $l<20$,
where we expect the Sachs-Wolfe effect to be the dominant effect, and we obtained, for
$l<20$, the angular power spectrum $l(l+1)C_l$ is divergent; clearly not consistent with
the observations from the CMB.

Finally, we showed the generic form that a primordial curvature power spectrum should
exhibit in the $\Rh$ framework to be consistent with the CMB temperature anisotropies
observed. Neither a scalar field dominating the early Universe satisfying an equation of
state $P(\phi) = -\rho(\phi)/3$ nor the inflaton yield a primordial spectrum compatible
with this requirement. Based on the results obtained in this paper,
we conclude that (in addition to the criticisms already raised by other
authors) it is not clear how to characterize the quantum perturbations within the
$R_h=ct$ Universe, rendering this model a very unlikely alternative to the standard
$\Lambda$CDM model.

\begin{acknowledgements}

G. R. B. acknowledges support from CONICET (Argentina) grant PIP
112-2012-0100540. G. L.'s research was funded by Consejo Nacional de Ciencia y
Tecnolog\'ia, CONACYT (Mexico). We thank the anonymous referee for useful suggestions that help to improve the presentation of our results and the overall clarity of the manuscript.

\end{acknowledgements}

\bibliographystyle{spphys}       
\bibliography{bibliografia_fulvio}   



\end{document}